\RequirePackage{rotating}
\documentclass[a4paper,fleqn,usenatbib,useAMS]{mnras}

\usepackage{graphicx}	
\usepackage{amsmath}	
\usepackage{amssymb}	
\usepackage{multicol}        
\usepackage{bm}		
\usepackage{pdflscape}	
\usepackage{rotating}   
\usepackage{esvect}
\usepackage[flushleft]{threeparttable}
\usepackage{booktabs,caption}

\usepackage[T1]{fontenc}
\usepackage{ae,aecompl}

\title[On the X-ray temperature of hot gas in diffuse nebulae]{On the
  X-ray temperature of hot gas in diffuse nebulae}
\author[Toal\'{a} \& Arthur]{J.A.\,Toal\'{a}\thanks{E-mail: j.toala@irya.unam.mx} and S.J.\,Arthur\\
Instituto de Radioastronom\'{i}a y Astrof\'{i}sica, UNAM Campus Morelia, Apartado postal 3-72, 58090 Morelia, Mich., Mexico}

\pubyear{}

\begin{document}
\label{firstpage}
\pagerange{\pageref{firstpage}--\pageref{lastpage}}
\maketitle

\begin{abstract}
X-ray emitting diffuse nebulae around hot stars are observed to have
soft-band temperatures in the narrow range [1--3]$\times10^{6}$~K,
independent of the stellar wind parameters and the evolutionary stage
of the central star. We discuss the origin of this X-ray temperature
for planetary nebulae (PNe), Wolf-Rayet nebulae (WR) and interstellar
wind bubbles around hot young stars in our Galaxy and the Magellanic
Clouds. We calculate the differential emission measure (DEM)
distributions as a function of temperature from previously published
simulations and combine these with the X-ray emission coefficient for
the 0.3--2.0~keV band to estimate the X-ray temperatures. We find that
all simulated nebulae have DEM distributions with steep negative
slopes, which is due to turbulent mixing at the interface between the
hot shocked stellar wind and the warm photoionized gas. Sharply peaked
emission coefficients act as temperature filters and emphasize the
contribution of gas with temperatures close to the peak position,
which coincides with the observed X-ray temperatures for the chemical
abundance sets we consider. Higher metallicity nebulae have lower
temperature and higher luminosity X-ray emission. We show that the
second temperature component found from spectral fitting to X-ray
observations of WR nebulae is due to a significant contribution from
the hot shocked stellar wind, while the lower temperature principal
component is dominated by nebular gas. We suggest that turbulent
mixing layers are the origin of the soft X-ray emission in the
majority of diffuse nebulae.

\end{abstract}




\section{INTRODUCTION}
\label{sec:intro}

Observations of the interstellar medium (ISM) with X-ray telescopes
have invariably shown the X-ray-emitting diffuse gas around hot stars
to have apparent temperatures in the narrow range
$T_\mathrm{X}\approx[1-3]\times$10$^{6}$~K. This applies to objects as
diverse as planetary nebulae (PNe) around low-mass stars
\citep[e.g.,][]{Kastner2012,Freeman2014}, Wolf-Rayet nebulae (WR)
around evolved massive stars \citep[e.g.,][and references
  therein]{Toala2017}, stellar wind bubbles around high-mass
main-sequence stars \citep[e.g.,][]{Gudel2008,Townsley2003}, and hot
gas in young star cluster environments \citep[e.g.,][]{Townsley2011a}.

Hot stars possess line-driven winds, which achieve highly supersonic
velocites with respect to their surrounding medium. The fast wind
forms an inward-facing shock when it interacts with the circumstellar
gas, which itself is swept up and shocked by an outward-facing shock
wave. The result is a hot, diffuse bubble of shocked stellar wind gas
surrounded by a dense shell of swept-up circumstellar material. Theory
predicts that the temperature in a hot, shocked stellar wind bubble
should depend principally on the stellar wind velocity $v_{\infty}$ in
the form \citep[see, e.g.,][]{Dyson1997}:
\begin{equation}
  k_\mathrm{B} T = \frac{3}{16}\mu m_\mathrm{H} v_{\infty}^{2},
\end{equation}
\noindent where $k_\mathrm{B}$, $\mu$, and $m_\mathrm{H}$ are the
Boltzmann constant, the mean particle mass and the hydrogen mass,
respectively. Accordingly, the range of observed stellar wind
velocities: 500--4,000~km~s$^{-1}$ for central stars of PNe
\citep[e.g.,][]{GuerreroDeMarco2013} and 600--3000~km~s$^{-1}$ for WR
stars \citep[e.g.,][]{Hamann2006,Hainich2015} should lead to orders of
magnitude variations in the X-ray temperatures and, moreover, the
expected temperatures are all in excess of $10^7$~K.

This discrepancy between theory and observations has been known for
quite some time \citep[see][]{Weaver1977} and various mechanisms have
been proposed to explain the lower-than-expected gas temperature in
the hot bubble.  Chief of these is the idea of thermal conduction
whereby heat diffuses from the hot, diffuse bubble into the
surrounding cooler, dense swept-up shell, carried by hot
electrons. The hot bubble loses heat to the surrounding shell, and the
inner surface of the dense shell evaporates into the bubble, raising
the density in the interface region. This scenario has been explored
analytically and numerically in the context of main-sequence stellar
wind bubbles \citep{Weaver1977,ReyesIturbide2009}, PNe
\citep{Soker1994,Steffen2008,ToalaArthur2016} and WR nebulae
\citep{Toala2011}. Although these models with thermal conduction can
produce gas at a few million degrees, the X-ray luminosity for
long-lived objects such as main-sequence stellar wind bubbles is
orders of magnitude higher than observations suggest. Better agreement
is found for the luminosities of short-lived objects such as PNe and
WR nebulae. In addition, there are several arguments against the
ubiquity of thermal conduction, the principal one being that the
presence of even a very small magnetic field will inhibit the
diffusion of the hot electrons.

\citet{Strickland1998} presented the first synthetic X-ray
observations obtained from 2D axisymmetric constant-wind purely
hydrodynamical numerical simulations of massive hot stars in a uniform
medium. Their X-ray spectra were convolved with the \textit{ROSAT}
calibration matrices to enable spectral fitting, in order to mimic
observed X-ray spectra. This work showed that even though spectral
fitting with one- and two-temperature emission plasma models can
statistically model the synthetic spectra, there is little indication
of the true nature of the physical properties within the hot bubble,
notably, the wide range of temperatures present ($10^4$ to $10^8$~K)
in the simulation. In particular, \citet{Strickland1998} found that
single temperature spectral fits lead to underestimates of the plasma
density and pressure in the X-ray emitting gas, which, in turn,
results in an underestimate in the true thermal energy in the
bubble. This is because single-temperature models are dominated by the
cool gas, which occupies only a small fraction of the total volume,
while the thermal energy is mostly contained in the hot gas, which
dominates the volume distribution. They concluded that the results of
simple spectral fits depend strongly on the instrument spectral
response and the spectral distribution of the source.

In a recent paper \citep[][hereafter Paper~I]{ToalaArthur2016} we
calculated the X-ray spectra and luminosities that would be produced
by the hot gas in a series of axisymmetric radiation-hydrodynamic
simulations of PNe, both with and without isotropic thermal
conduction. Over the energy range of corresponding to the
\textit{Chandra} telescope soft band we found that the
emission-coefficient-weighted mean temperature of all the simulations
as a function of time was remarkably constant in the temperature range
reported by observations (1--3$\times$10$^{6}$~K), even though each
simulation contained gas at all temperatures in the X-ray emitting
range between $10^5$ and $10^8$~K. We suggested that this could be an
explanation for the results of single-temperature spectral fits.

In the present paper, we explore the properties of the emission
coefficient calculated for a variety of metallicities and the soft
X-ray bands (from {\it Chandra} and {\it XMM-Newton}) and how it
affects the interpretation of the hot, diffuse gas temperature in PNe
and WR nebulae around evolved stars and also stellar wind bubbles
around hot young stars.

\section{METHODS}
\begin{figure*}
\begin{center}
\includegraphics[angle=0,width=.33\linewidth]{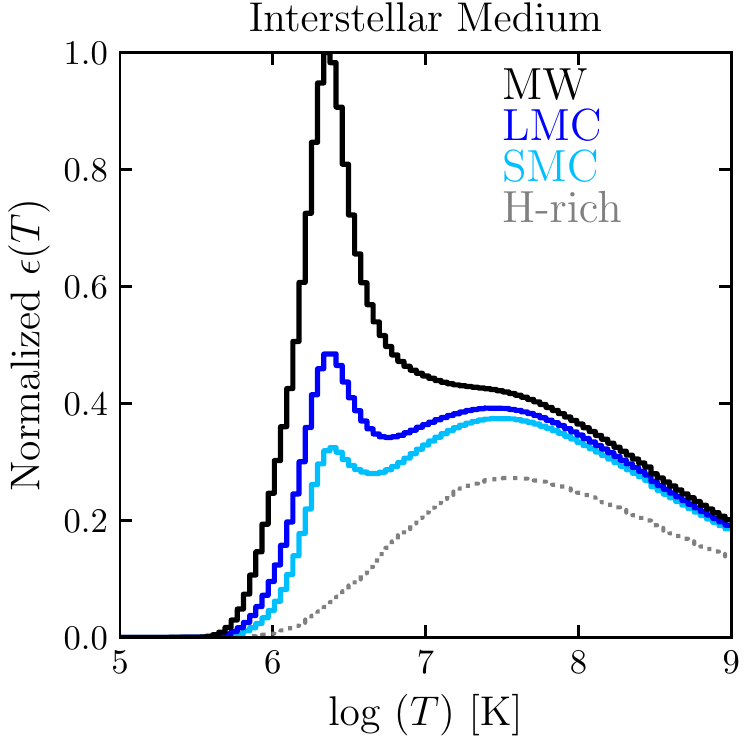}~
\includegraphics[angle=0,width=.33\linewidth]{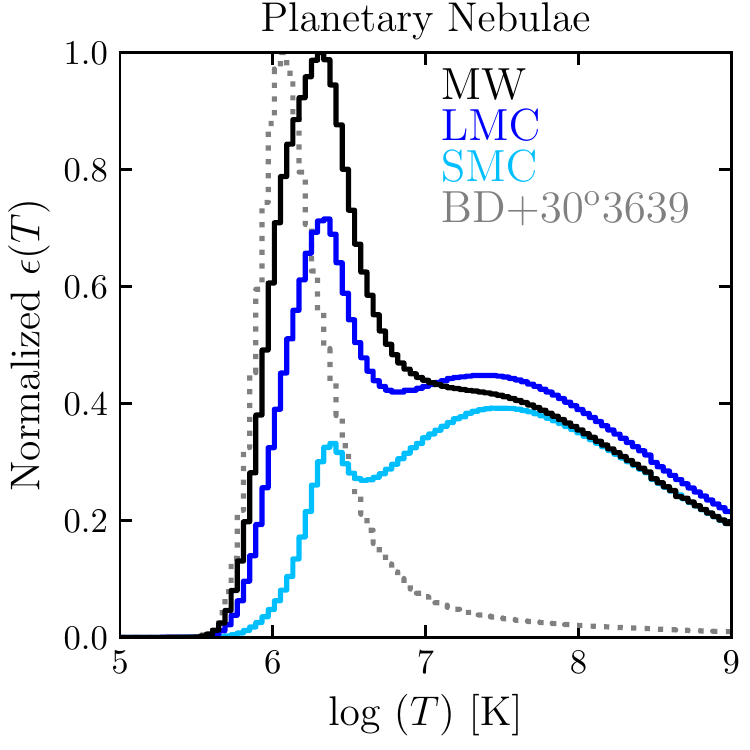}~
\includegraphics[angle=0,width=.33\linewidth]{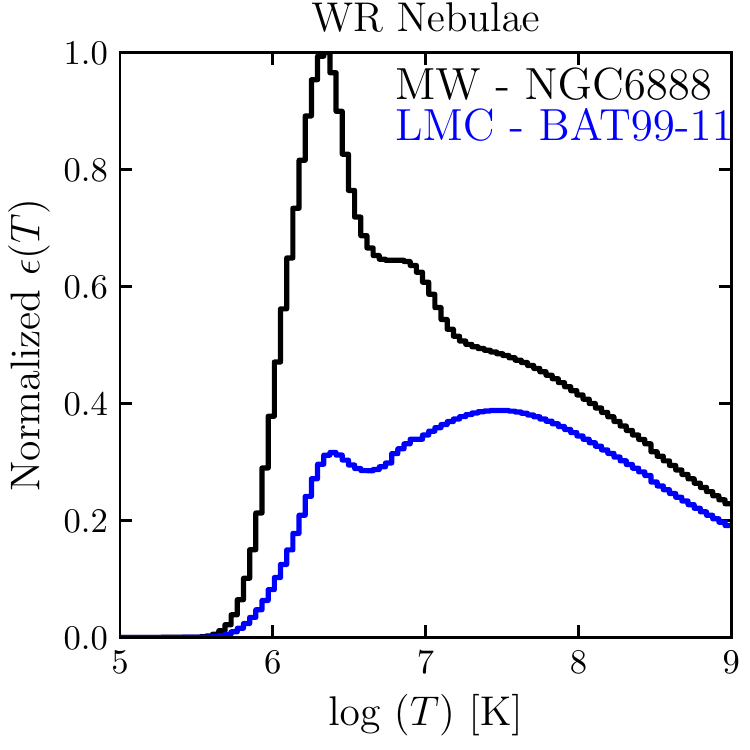}
\end{center}
\caption{Emission coefficient $\epsilon(T)$ for the 0.3--2.0~keV
  energy range computed for different abundance sets. Left: ISM
  abundances from the Milky Way (MW) and the Large and Small
  Magellanic Clouds (LMC and SMC). For comparison we also show the
  result for a pure hydrogen gas. Center: PNe abundances. Right:
  Computed for two iconic WR nebulae abundances. The central panel
  also shows the results for the hydrogen-poor PN
  BD$+$30$^{\circ}$3639 abundance set, which has been divided by 10 to
  fit on the same scale. The emission coefficient in each panel has
  been scaled to the corresponding MW value.}
\label{fig:temp_par}
\end{figure*}

We begin by defining the two main tools we use in our analysis: the
emission coefficient and the differential emission measure.

\subsection{Emission coefficient}

\begin{figure*}
\begin{center}
  \includegraphics[angle=0,width=0.45\linewidth]{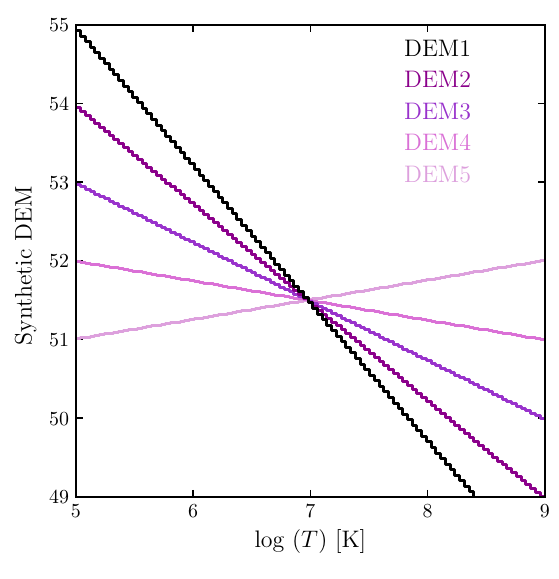}~
  \includegraphics[angle=0,width=0.45\linewidth]{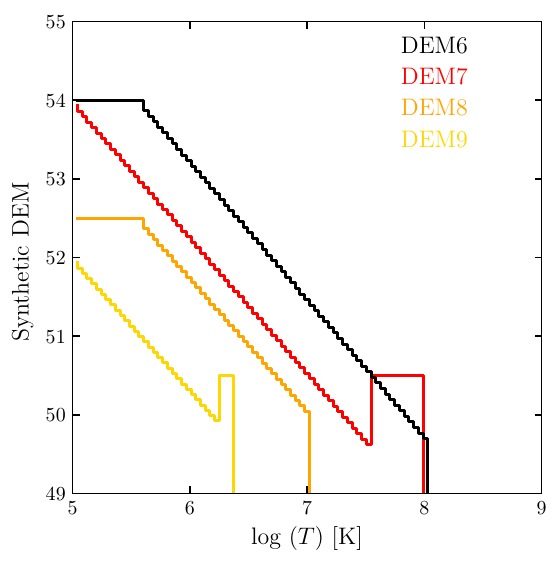}~
\end{center}
\caption{Representative synthetic differential emission measure
  (DEM) profiles. Left panel: generic DEM profiles with different
  slopes. Right panel: profiles typical of wind bubbles with
  evolving central stars. }
\label{fig:dem_syn}
\end{figure*}

The emissivity $\epsilon(T,E)$ takes into account the sum of all the
contributions to the emission spectrum (e.g., collisionally excited
lines, free-free, free-bound and two-photon continua, etc.) made by
diffuse gas at a single temperature $T$. It also depends on the
chemical abundances.  The individual spectra for 100 temperature bins
at 0.04~dex intervals in the range $5 \le \log(T) \le 9$ were
calculated using the {\sc chianti} atomic database\footnote{{\sc
    chianti} is an extensively tested database and software package
  \citep{Dere1997,Dere2009,Landi2013} tailored to study the UV and
  X-ray emission from hot plasmas.}. The spectra all have bin widths
of 0.01~\AA\, and a spectral-line full width at half maximum of
0.1~\AA\, is assumed. The emission coefficient over the X-ray energy
band $[E_1,E_2]$ is then defined as
\begin{equation}
  \epsilon(T) = \int^{E_\mathrm{2}}_{E_\mathrm{1}}{\epsilon(T,E)dE} \
  \ .
\end{equation}
For the X-ray telescopes \textit{Chandra} and \textit{XMM-Newton}, the
soft energy band corresponds to $E_\mathrm{1}=0.3$~keV and $E_\mathrm{2}=2.0$~keV.

In Figure~\ref{fig:temp_par} we plot the emission coefficient
calculated for the following different abundance sets (see Appendix A
for actual values):
\begin{enumerate}
\item Interstellar medium (ISM) abundances in the Milky Way (MW), the
  Large Magellanic Cloud (LMC) and the Small Magellanic Cloud (SMC).
\item Planetary nebulae (PN) abundances in the MW, LMC and SMC. Also,
  the hydrogen-poor PN BD$+30^{\circ}3639$, which is the brightest PN
  in X-rays in our Galaxy.
\item The representative WR nebulae NGC~6888, which is the most
  studied WR nebula in the MW, and BAT99-11 in the LMC.
\item Pure hydrogen gas.
\end{enumerate}
For all the abundance sets except the pure hydrogen case, the emission
coefficient has two characteristic features: a narrow peak at
1--$3\times 10^6$~K and a broad plateau or bump at higher
temperatures. The exact position of the narrow peak, and the relative
heights of the peak and broad bump change with metallicity. For Milky
Way ISM abundances the narrow peak is twice the height of the broad
bump, while for SMC ISM abundances the broad bump is marginally higher
than the narrow peak. The pure hydrogen case has only the broad,
high-temperature feature, thus the narrow peak can be associated
directly to line emission from metals.

\subsection{Differential Emission Measure}

A useful tool for summarizing the hot gas properties of a numerical
simulation, under the assumption that the hot bubble is optically thin
at X-ray wavelengths, is the Differential Emission Measure (DEM),
which we define as
\begin{equation}
\mathrm{DEM}(T_\mathrm{b}) = \sum_{k, T_k \in T_\mathrm{b}} n_\mathrm{e}^2 \Delta V_k,
\label{eq:DEM}
\end{equation}
where $n_\mathrm{e}$ is the
  electron number density in cell $k$, 
  $\Delta V_k$ is the volume of cell $k$ and the sum is performed over
  cells with gas temperature falling in the bin whose central
  temperature is $T_\mathrm{b}$ (see also \citealp{Strickland1998}).

The DEM profile will be different for different types of astrophysical
object and, for a given object, will evolve with time. For example, a
PN has a stellar wind whose terminal velocity increases rapidly over a
short timescale. This means that the temperature behind the inner wind
shock will increase with time.

In Figure~\ref{fig:dem_syn} we show synthetic DEMs, which cover a
range of hypothetical behaviours. The left panel, showing DEM1--5,
presents arbitrary DEMs that illustrate a variety of different slopes
corresponding to power-law indices from $-1.75$ to $+0.25$. The right
panel, showing DEM6--9, presents DEMs that are characteristic of
stellar wind bubbles, WR nebulae and PNe such as those we have modeled
using numerial simulations in previous work \citep[see][and
  Paper~I]{Toala2011}. The temperature spread of these profiles
reflects both the evolution of the stellar wind velocity and the
effect of mixing seen at the edge of the hot bubble in our 2D
simulations. The profiles with flat regions at $T < 10^6$~K mimic the
behaviour of models with thermal conduction, where the lower
temperature region corresponds to the conduction layer. Note that a
uniform velocity stellar wind with no mixing
\citep[e.g.,][]{Arthur2012} would give a DEM profile consisting of a
single spike in the temperature bin corresponding to the postshock
temperature.

\subsection{Mean temperature}
\label{sec:meantemp}

X-ray observations of hot bubbles often report a single 
plasma temperature ($T_\mathrm{X}$). This plasma temperature is
obtained by fitting a single-component optically-thin plasma emission model to
the observed spectrum (\textit{apec} or \textit{mekal} model).

We can use the emission coefficient and the DEM distribution to
calculate an average temperature for the hot bubbles generated by
numerical simulations. This temperature will be closer to the
observationally derived X-ray temperature than either a mass-weighted
or volume-weighted mean temperature \citep[see, e.g.,][]{Rogers2014}.
\citet{Strickland1998} defined an intrinsic (i.e., unabsorbed)
flux-weighted average temperature $\langle T_\mathrm{EW} \rangle$ over
the whole 0.005--15~keV energy range. Their definition results in very
low average temperatures because it gives too much weight to cool gas
emitting mainly in the UV.

In Paper~I we defined the mean temperature as
\begin{equation}
  T_\mathrm{A} = \frac{\int{\epsilon(T) \mathrm{DEM}(T) T dT}}{\int{\epsilon(T) \mathrm{DEM}(T) dT}},
\label{eq:avtemp}
\end{equation}
where $\epsilon(T)$ is the emission coefficient in the X-ray band and
$\mathrm{DEM}(T)$ is the differential emission measure at temperature
$T$. The integral is performed over all the temperature bins of the
simulation. This mean temperature is weighted by both emission
coefficent and DEM and so takes into account the chemical abundances
together with the mass and volume distribution of the hot gas.
Although the $\mathrm{DEM}(T)$ profile itself may be weighted towards
low temperatures ($T\sim10^{5}$~K), the fact that the emission
coefficient $\epsilon(T)$ is sharply peaked around a few times
$10^6$~K means that gas with this temperature is preferentially
selected (see figure~5 in Paper~I), i.e., the emission coefficient
acts as a temperature filter.

\begin{table*}
 \centering
 \caption{Average temperatures ($T_\mathrm{A}$) in logarithmic values calculated for different DEMs and abundance sets.}
 \label{tab:averaged_temp}
 \begin{tabular}{ccccccccccc}
  \hline
  Abundance & DEM1  & DEM2 & DEM3 & DEM4 & DEM5 & DEM6 & DEM7 & DEM8 & DEM9 & DEM$_\mathrm{f}$ \\
  set       &       &      &      &      &      &      &      &      &      &                 \\
  \hline
  ISM-MW    & 6.306  & 6.502& 6.928 &  7.614& 8.237  &6.300 & 6.482 & 6.268 & 6.218 & 7.957\\
  ISM-LMC   & 6.384  & 6.630& 7.107 &  7.757& 8.293  &6.374 & 6.644 & 6.314 & 6.230 & 8.055\\
  ISM-SMC   & 6.459  & 6.739& 7.216 &  7.826& 8.315  &6.446 & 6.774 & 6.362 & 6.243 & 8.098\\
  \hline
  PN-MW     &6.200  &6.379  &6.802  &7.529 & 8.210  &6.197 & 6.335 & 6.176 & 6.168 & 7.903\\
  PN-LMC    &6.262  &6.481  &6.955  &7.662 & 8.264  &6.255 & 6.457 & 6.218 & 6.189 & 7.998\\
  PN-SMC    &6.461  &6.742  &7.230  &7.837 & 8.319  &6.448 & 6.784 & 6.360 & 6.242 & 8.105\\
  BD$+30^{\circ}$3639 &6.060 & 6.127  &6.263 & 6.696  &7.573 & 6.060 & 6.076 & 6.062 & 6.100 & 7.102\\
  \hline
  WR-MW (NGC\,6888) & 6.281  &6.478  &6.905  &7.593 & 8.226 &6.276 & 6.448 & 6.245 & 6.202 & 7.940\\
  WR-LMC (BAT99-11) & 6.400  &6.685  &7.186  &7.814 & 8.312 &6.388 & 6.703 & 6.310 & 6.216 & 8.092\\
  \hline
 \end{tabular}
\end{table*}

\subsection{Test Results}
\label{sec:testres}

Table~\ref{tab:averaged_temp} reports the average temperatures
obtained from combining the emission coefficients of
Figure~\ref{fig:temp_par} with the DEM distributions of
Figure~\ref{fig:dem_syn} using Equation~\ref{eq:avtemp}. Also included
is the mean temperature that would be obtained from a completely flat
DEM distribution (column DEM$_\mathrm{f}$). There are two clear
trends. Firstly, the mean temperature increases as the DEM slope (as a
function of bin temperature) goes from steeply negative to
positive. Secondly, the mean temperature increases with decreasing
metallicity. Thus, for example, objects with a similar history (same
DEM) will appear to be hotter in the SMC than in the Milky Way. Also,
average temperatures for objects with enhanced abundances such as PNe
and WR nebulae will be lower than objects such as stellar wind bubbles
with ISM abundances, even though the DEM profiles are similar.

\section{Results}
We apply our methods to the results of 2D axisymmetric numerical
simulations of PNe, WR nebulae and stellar wind bubbles in HII
regions. These simulations have been reported in
\citet{ToalaArthur2014}, Paper~I, \citet{Toala2011} and
\citet{Arthur2006}, respectively. Full details of the numerical codes
used can be found in the cited references. In particular, in
simulations with thermal conduction, the conduction is isotropic and
saturation is taken into account by limiting the electron mean free
path \citep{ToalaArthur2014}.

\subsection{Simulated PNe}

\begin{figure}
\begin{center}
\includegraphics[angle=0,width=0.93\linewidth]{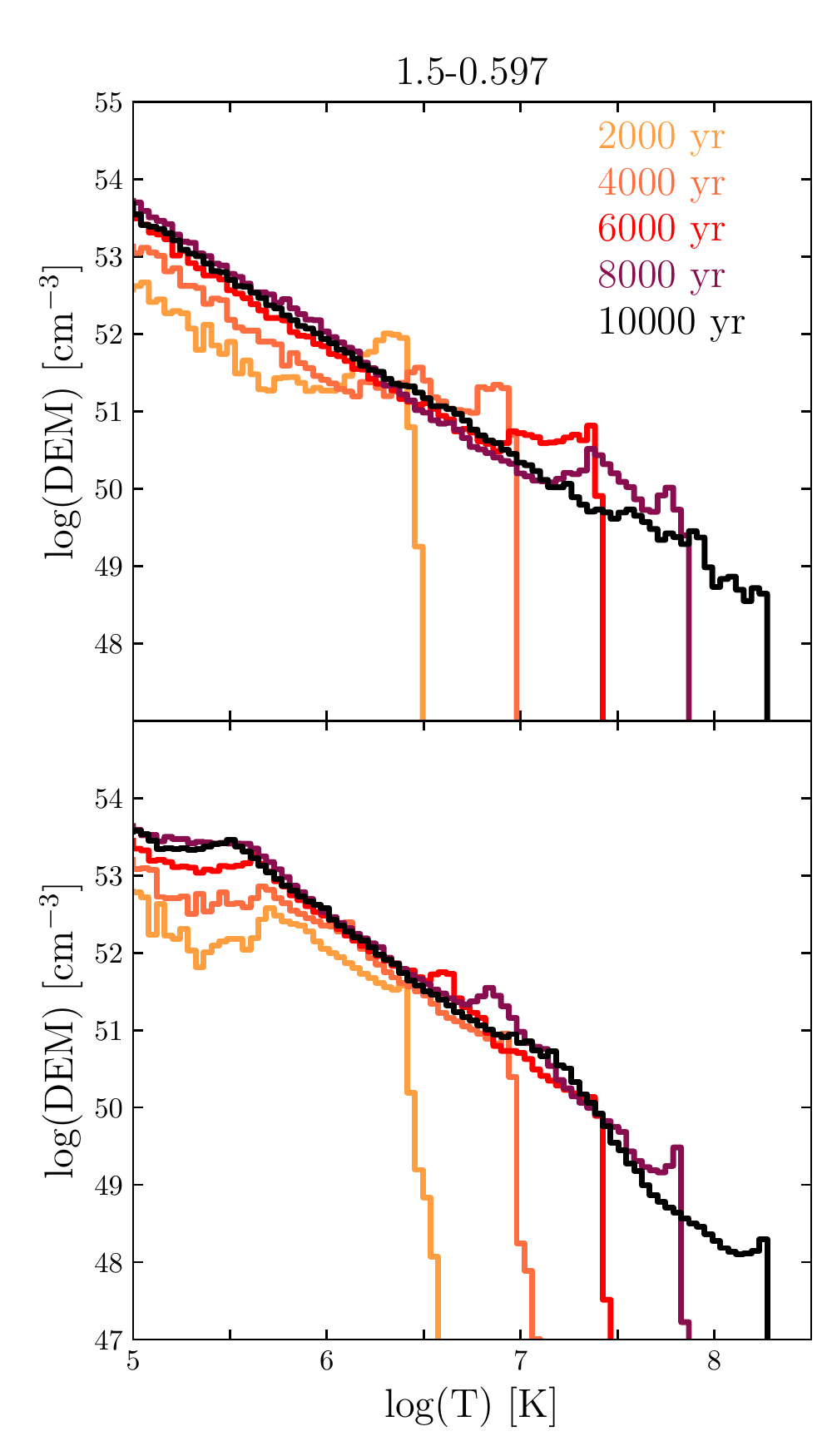}
\end{center}
\caption{DEM profiles from selected times in the evolution of a PNe
  around a central star whose initial ZAMS mass was $1.5 M_\odot$ and
  whose final white dwarf mass will be $0.597 M_\odot$ (see Paper~I).}
\label{fig:dem-pn15}
\end{figure}

For PNe, the timescales are so short (less than 10,000~yr) that
radiative cooling is not yet important for the hot, shocked fast wind
gas. The DEM profiles for a set of simulations corresponding to a $1.5
M_\odot$ progenitor star that ends its life as a $0.597 M_\odot$ white
dwarf are shown in Figure~\ref{fig:dem-pn15}. The upper panel is for
simulations without thermal conduction, while the lower panel is for
simulations where thermal conduction was included. Both sets of
results show that the DEM profiles become steeper with time and that
the maximum temperature of the profile moves to higher values. This is
because the stellar wind velocity of the central star increases with
time and the maximum temperature reflects the immediate postshock
temperature at the inner wind shock.

The continuous spread of the DEM profile with temperature is typical
of wind bubbles with turbulent mixing layers. Instabilities in the
swept-up shell lead to the formation of filaments and clumps that
interact with the fast wind and the ionizing photons resulting in gas
with intermediate densities and temperatures between the hot, shocked
fast wind and the warm, photoionized dense shell. The nebular gas is
heated by shocks formed in hydrodynamic interactions in material
photoevaporated or ablated from the filaments and clumps. Pressure
fluctuations in the shocked fast wind as it squeezes past the
obstacles lead to temperature fluctuations in the hot plasma. These
hydrodynamic effects are not seen in 1D simulations (see Paper I).

\begin{figure}
\begin{center}
\includegraphics[angle=0,width=\linewidth]{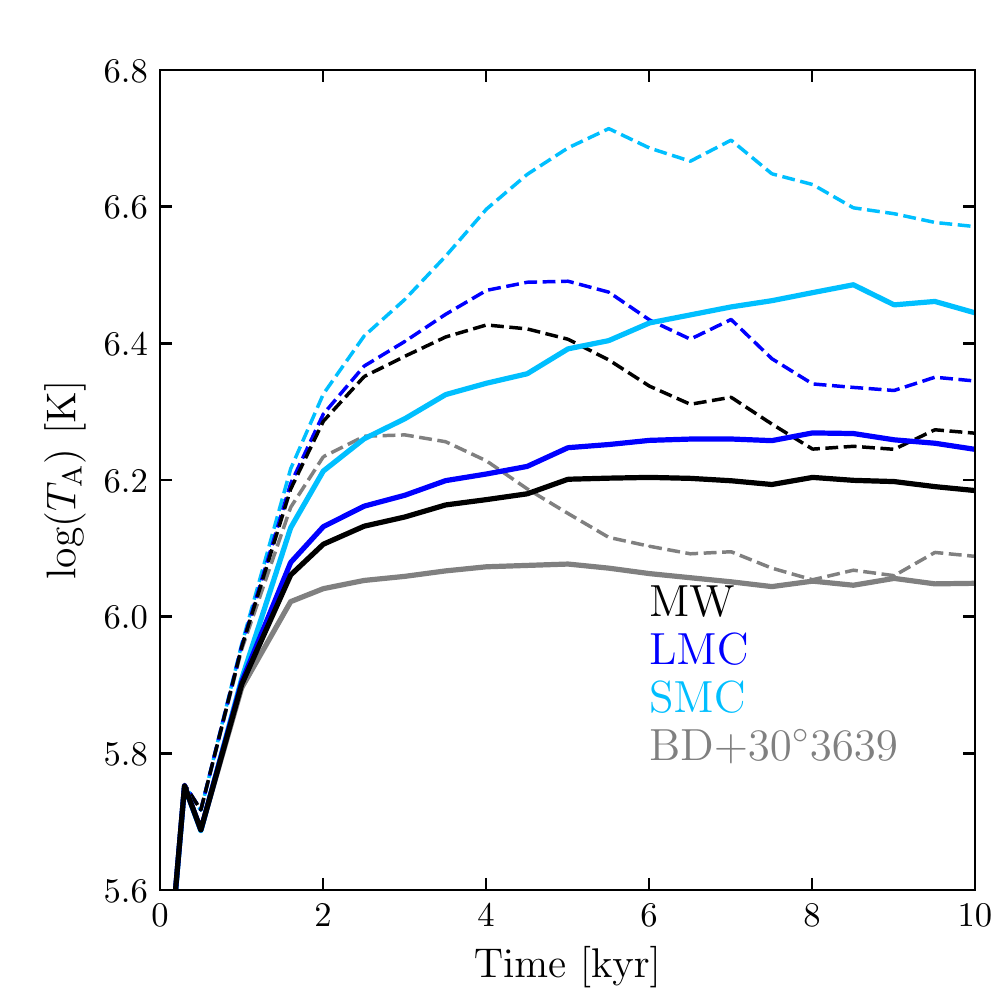}
\end{center}
\caption{Evolution of the emission-coefficient-weighted mean
  temperature of simulated PNe for a $1.5 M_\odot$ progenitor star for
  different abundance sets. The solid lines correspond to
  simulations with thermal conduction, the dashed lines are for
  simulations without conduction.}
\label{fig:pntasim}
\end{figure}

When conduction is included in the calculations, thermal energy is
efficiently transferred from the hottest gas to the dense shell where
it evaporates material into the conduction layer. This smoothes out
the DEM profile at higher temperatures and the conduction layer
results in the characteristic plateau for $\log_\mathrm{10} T < 5.7$.
The main body of the DEM profile has a very similar slope to the
models without conduction but the DEM values are higher for a given
temperature. 

The emission-coefficient-weighted mean temperature as a function of
time for these simulations is plotted in Figure~\ref{fig:pntasim} for
PNe abundances in the Galaxy, the LMC and the SMC, and can be
understood in the context of the test cases described in
Section~\ref{sec:testres}. The slopes of the DEM profiles for the
models without conduction steepen with time and so we would expect the
average temperature at early times to be higher but then tend to the
peak value of the corresponding emission coefficient at later
times. For the models with conduction, the plateau at $\log_{10} T <
5.7$~K and smooth slope all the way to the maximum temperature lead to
faster convergence of the mean temperature. The abundance set
determines the maximum mean temperature: the Milky Way abundances
result in $\log_{10} T_\mathrm{A} < 6.4$ and a converged value
$\log_{10} T_\mathrm{A} \sim 6.2$, while the SMC abundances have
$\log_{10} T_\mathrm{A} < 6.7$, which does not converge over the
10,000~yr timescale because of the competition between the DEM profile
slope and the relative heights of the narrow and broad peaks of the
emission coefficient (see Fig.~\ref{fig:temp_par}).

\begin{figure}
\begin{center}
\includegraphics[angle=0,width=\linewidth]{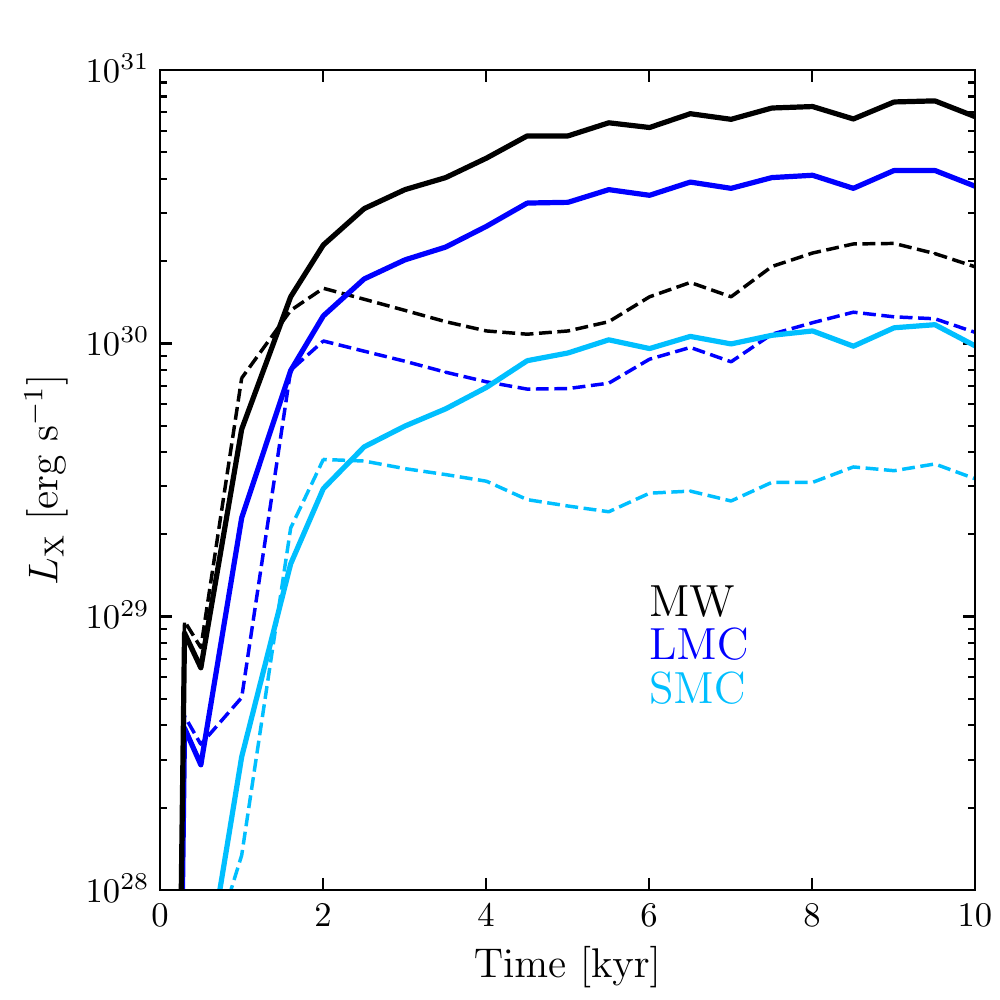}
\end{center}
\caption{Evolution of the X-ray luminosity of simulated PNe for a $1.5
  M_\odot$ progenitor star for different abundance sets. The solid
  lines correspond to simulations with thermal conduction, the dashed
  lines are for simulations without conduction.}
\label{fig:pnlumsim}
\end{figure}

Figure~\ref{fig:pnlumsim} shows the luminosities corresponding to the
same models as Figure~\ref{fig:pntasim}. Results for the models with
and without thermal conduction are presented. After a short initial
evolution, the luminosities remain roughly constant for both types of
model. It can be seen that lower metallicity objects are much less
luminous, which has consequences for the possible detection of PNe in
external, low-metallicity galaxies.

\subsection{Simulated Wolf-Rayet Nebulae}

WR nebulae are similar to PNe, in the sense that a fast, radiatively
driven wind from a hot compact object interacts with a circumstellar
medium formed by previous intense mass-loss episodes of the progenitor
star.  On the other hand, WR nebulae are different to PNe in that the
mechanical luminosity of the wind does not drop off with time and the
timescale of this evolutionary stage is much longer. Thus, the shocked
wind material in the nebula becomes proportionately more significant
as time progresses.  There will be chemical abundance differences
between the hydrogen-rich nebular gas and the fast wind material
leaving the hydrogen-poor surface of the star. In particular, all
elements such as oxygen, silicon, magnesium and iron should be in the
gas phase in the stellar wind material, whereas in the nebular gas
they may be partially locked up in silicate dust grains. Higher
energy diffuse X-ray emission should reflect the shocked stellar wind,
while the lower energy part of the spectrum will be produced by the
nebular material.

The evolution of massive stars prior to the WR stage is very varied,
depending on progenitor initial mass, metallicity and stellar rotation
rate \citep[][]{Ekstrom2012,Eldridge2004,Eldridge2006}. The mass
expelled during the red supergiant (RSG) or luminous blue variable
(LBV) stage can be distributed close to the WR star or in several
shells at increasing distances, depending on the details of the
mass-loss history \citep{Toala2011}. Moreover, the velocity of the
circumstellar material will below ($\sim 30$~km~s$^{-1}$) for RSG
material but considerably higher ($\sim 200$~km~s$^{-1}$) for LBV
material. These factors affect the details of the wind-wind
interaction and the nature of the instabilities that form in the
interaction region.

\begin{figure}
\begin{center}
 \includegraphics[height=1.6\linewidth]{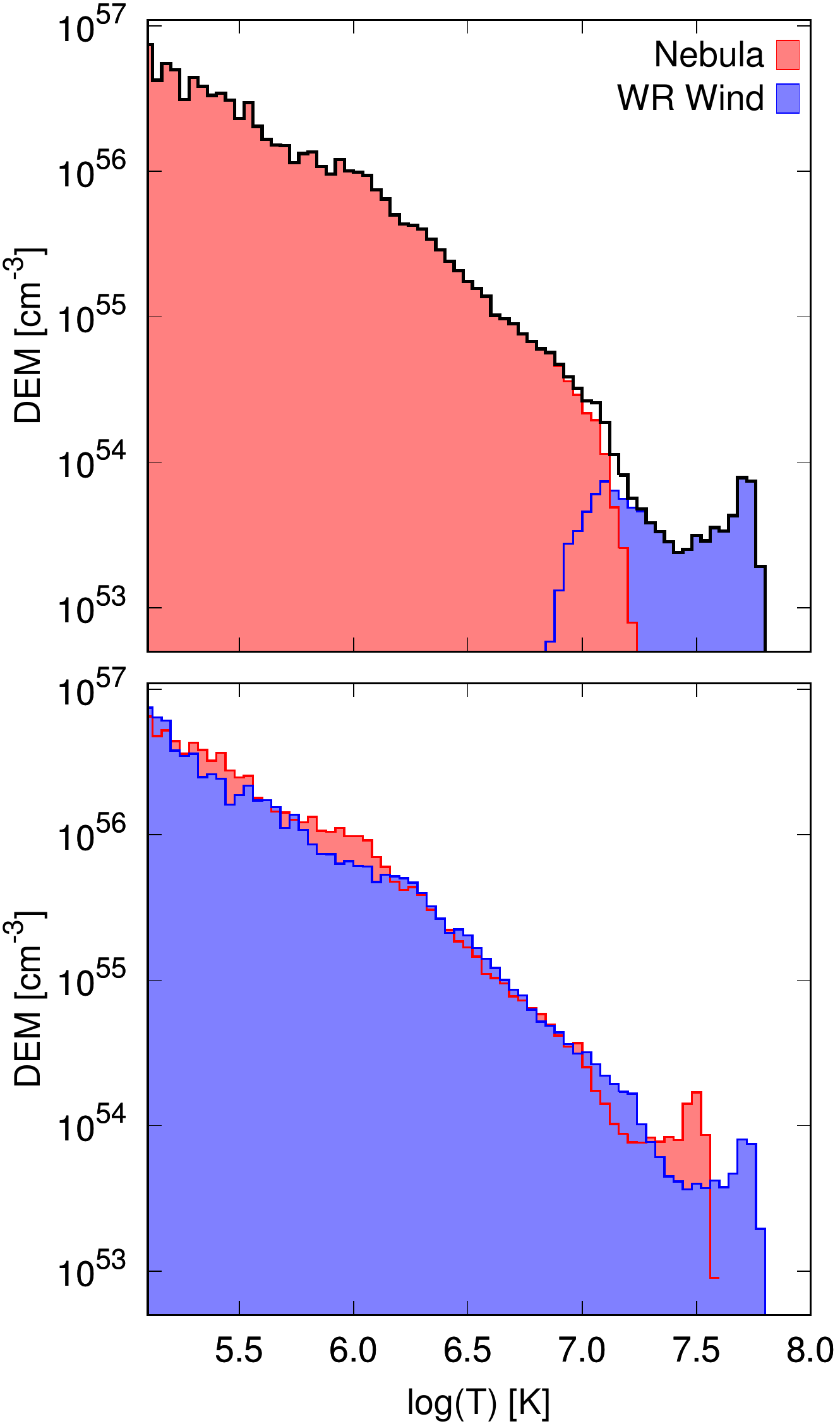}
\end{center}
\caption{DEM profile for the numerical simulation. Top panel: red
  histogram  is the nebular contribution (scalar~$> 0.5$) and the blue
  histogram is the WR fast stellar wind (scalar~$< 0.5$)
  contribution. Bottom panel: DEM profiles if the abundances were
  wholly nebular (red) or wholly WR star (blue) values.}
\label{fig:wrsimdem}
\end{figure}

In Appendix~\ref{sec:appwr} we show the temperature and nebular gas
(using an advected scalar label) distributions for a 2D axisymmetric
numerical simulation of a simplified WR nebula consisting of a
constant velocity fast wind ($v_{\infty}= 1600$~km~s$^{-1}$)
interacting with a circumstellar medium generated by a constant slow
dense wind that ejected $15 M_\odot$ of material over a timescale of
200,000~years. Figure~\ref{fig:wrsimdem} shows the DEM profile for
this simulation, depicted 15,000 years after the onset of the fast
wind. Note that the interaction region produces a full spectrum of gas
temperatures, with the nebular gas being heated up to $\log_{10} T
\sim 7.2$ by shocks refracted by the clumps and filaments in the
swept-up RSG material. On the other hand, the shocked fast-wind
temperatures extend down to $\log_{10} T \sim 7.2$ due to pressure and
density variations in the complex non-radial flows in the interaction
region. The separation in temperature of nebular and wind material is
maintained throughout the simulation.

\begin{figure}
\begin{center}
\includegraphics[angle=0,width=1\linewidth]{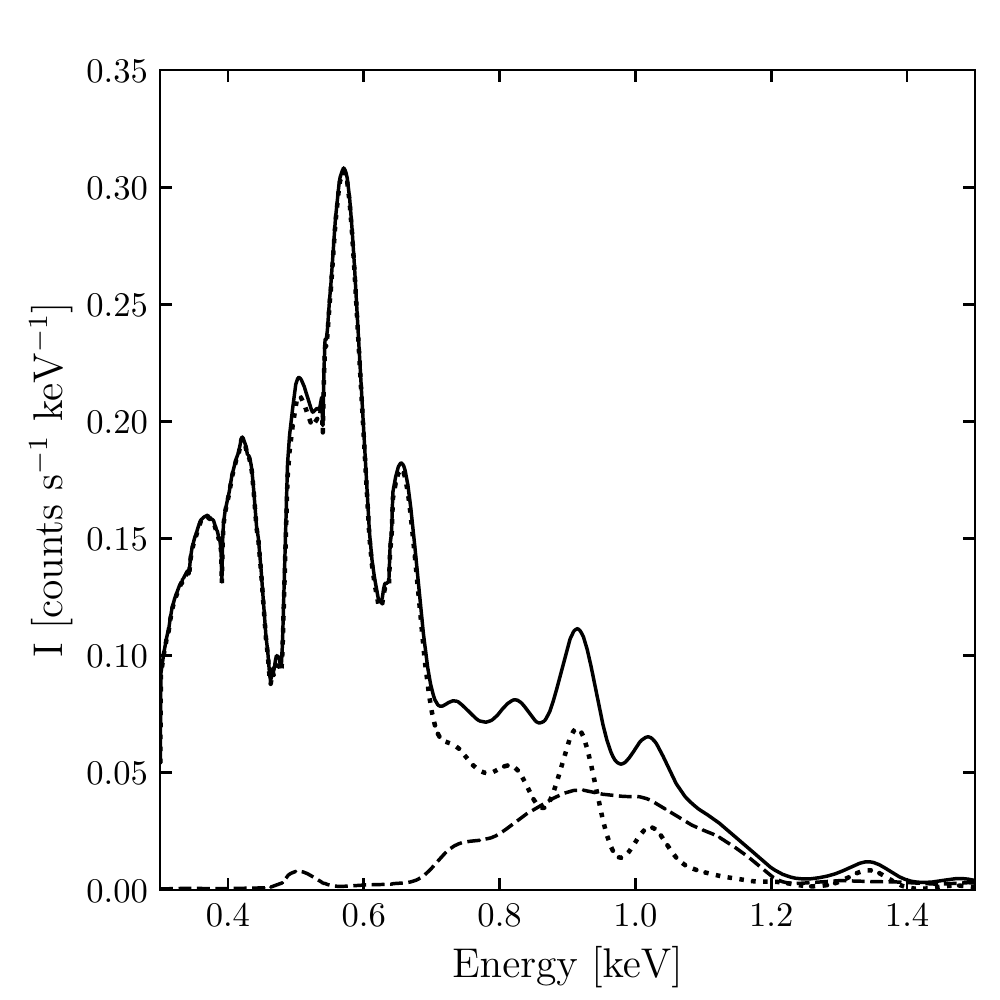}
\end{center}
\caption{Synthetic spectrum obtained from the DEM profiles shown in
  Figure~\protect\ref{fig:wrsimdem} (top panel). The spectrum has been
  corrected for absorption by a neutral hydrogen column density of
  $N_\mathrm{H}=3.13\times$10$^{21}$~cm$^{-2}$ and convolved with the
  instrumental response matrices of the \textit{Chandra} ACIS-S
  instrument. The spectrum is binned to a resolution of 60~eV. The
  solid line represents the total spectrum, the dotted line represents
  the contribution of the nebular gas and the dashed line shows the
  contribution of the hot, shocked fast-wind gas}.
\label{fig:wrspecsim}
\end{figure}

The spectrum obtained from the simulation DEM profile (see
Fig.~\ref{fig:wrsimdem}) assuming abundances as for the galactic WR
nebula NGC\,6888 (see Table~\ref{tab:abundances}) is shown in
Figure~\ref{fig:wrspecsim}. The intrinsic spectrum has been corrected
for the interstellar absorbing neutral hydrogen column density
reported for its central star WR\,136
\citep[$N_\mathrm{H}=3.16\times10^{21}$~cm$^{-2}$;][]{Hamann1994} and
convolved with the instrumental response matrices of \textit{Chandra}
ACIS-S. The total spectrum (solid line) and the contribution of the
shocked fast wind (dashed line) indicate that the WR fast wind
contributes considerably in the 0.8--1.0~keV spectral energy
range. The spectrum shows strong similarities with the observed
spectra reported by \citet{Toala2014} \citep[see also][for comparison
  with {\it XMM-Newton} spectra]{Toala2016a}: Most of the emission
comes from energies below 0.7~keV with significant emission from the
Fe complex and Ne\,{\sc ix} line at 0.8--1.2~keV, declining to
energies above 1.5~keV. As in other WR nebulae \citep[see,
  e.g.,][]{Toala2012}, the observed spectrum of NGC\,6888 was fit with
a two-temperature \textit{apec} model ($1.4\times10^6$~K and
$8.2\times10^6$~K). The second component is necessary to give a good
fit for energies above 0.8~keV. In our synthetic spectrum, the hot
shocked fast wind continuum emission contributes about half the counts
in this range and peaks are lines from an iron complex whose
prominance is due to the high iron abundance in the nebular gas, which
is closer to solar than to ISM or PNe values in this object.

\subsection{Simulated stellar wind bubbles in HII regions}

\begin{figure}
\begin{center}
 \includegraphics[height=1.\linewidth]{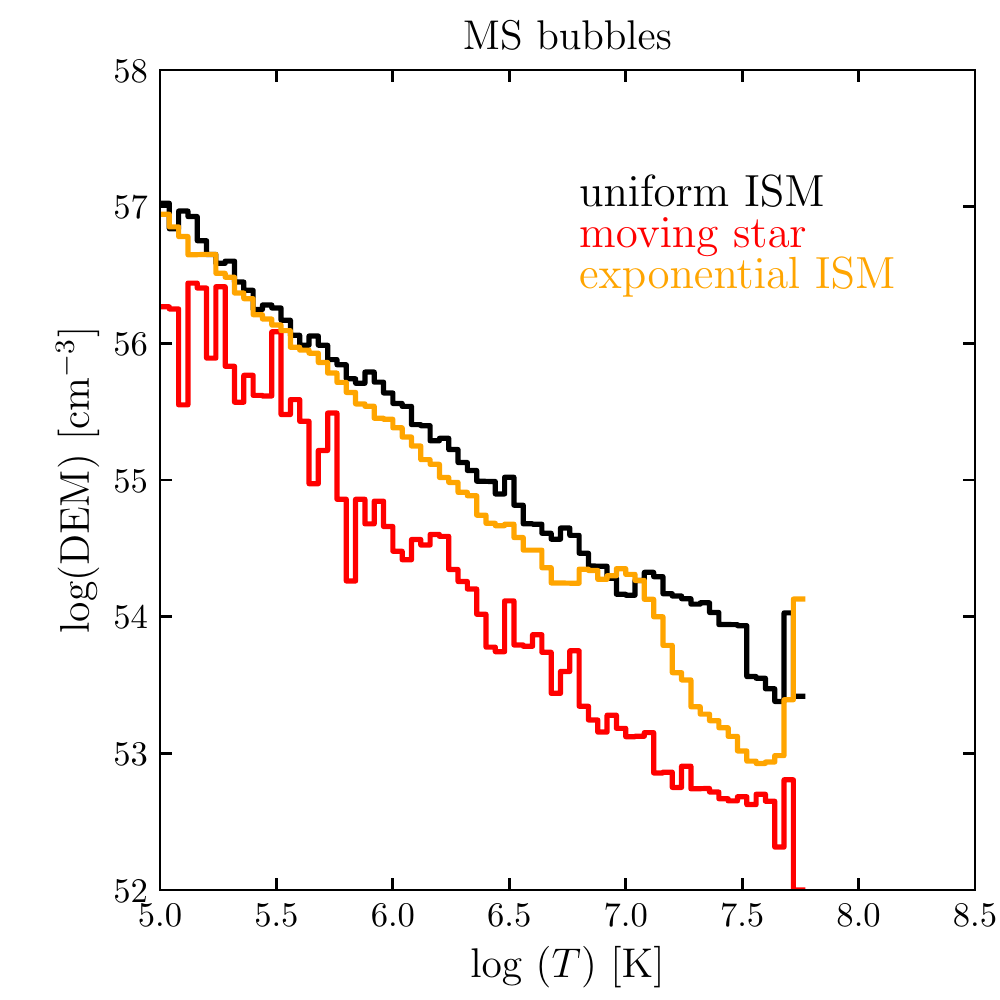}
\end{center}
\caption{DEM profiles for a stellar wind bubble in an HII region in a
  uniform medium (black line), an exponential density distribution
  (yellow line) and a star moving up a density gradient (red line). In
  each case the simulation represents 20,000~yrs of evolution and the
  stellar wind parameters are identical (see text).}
\label{fig:mssimdem}
\end{figure}

Following the work of \citet{Mackey2015}, who studied stellar wind
bubbles in HII regions around single O stars moving in a uniform
medium, we decided to re-examine the simulations of
\citet{Arthur2006}. The DEM profiles calculated for some of their
models are shown in Figure~\ref{fig:mssimdem}. The models that were
available correspond to a stellar wind bubble and HII region evolving
in (i) a uniform medium with density $n_0 = 6000$~cm$^{-3}$, (ii) a
plane-stratified medium with exponential density distribution
$n=n_0\exp(-z/H)$ where the scale height is $H=0.2$~pc and
$n_0=8000$~cm$^{-3}$ (Model E of \citealp{Arthur2006}) and (iii) a
star moving up a density gradient with velocity $V_* = 10$~km~s$^{-1}$
(Model H of \citealp{Arthur2006}). In each model the stellar wind
velocity is $v_{\infty}=2000$~km~s$^{-1}$ and the mass-loss rate is
$\dot{M} = 10^{-6} M_\odot$~yr$^{-1}$, while the ionizing photon rate
is $2.2\times 10^{48}$~s$^{-1}$.

What we see in each case is a steep, negative DEM profile, very
similar to the previous results seen for PNe and WR nebulae. Combining
these DEM distributions with Milky Way ISM abundances, we find
emission-coefficient-weighted mean temperatures of $\log(T_\mathrm{A})
= 6.40$, 6.38 and 6.32 for the uniform medium, exponential medium and
moving star models, respectively. This is to be expected, given the
form of the DEM profile.

\subsection{Temperature from Spectra}
In order to evaluate whether the averaged temperature ($T_\mathrm{A}$)
is a good estimate to what is obtained from observations
($T_\mathrm{X}$), that is, that $T_\mathrm{A} \approx T_\mathrm{X}$,
we took one of the synthetic spectra presented in Paper~I and used it
as input in {\sc xspec} \citep[v12.9.0;][]{Arnaud1996}. We used the
synthetic spectrum of the model 1.5-0.597 at 8000~yr without thermal
conduction convolved by the {\it Chandra} telescope matrices presented
in figure~17 in Paper~I which has log($T_\mathrm{A}$)=6.24
($kT_{A}$=0.151~keV). The spectrum was modelled assuming the same set
of abundances for PNe (see column~5 in Table~A1) with a fixed absorbing
column density $N_\mathrm{H}$=8$\times$10$^{20}$~cm$^{-2}$. The
best-fit model turned out to have a plasma temperature of
$kT$=0.140~keV, a difference of $\sim$8 per cent with its
corresponding $T_\mathrm{A}$. Thus, we are confident that the averaged
plasma temperature obtained with the method described in
Section~\ref{sec:meantemp} is a good estimate of the temperatures
obtained from observations. We remark that the single temperature
approach is not representative of the spatial distribution of
temperatures in the hot bubble.

\section{Discussion}

There are 3 main aspects to the temperature of soft X-rays in diffuse
objects: the chemical abundances, the actual distribution of
temperatures in the diffuse object (i.e., the DEM profile) and the
quality of the observations. We discuss PNe, WR nebulae, main sequence
wind bubbles and cluster environments in the context of these
considerations.

\subsection{Planetary Nebulae} 
The shapes of the emission coefficients shown in
Figure~\ref{fig:temp_par} are a direct result of the set of chemical
abundances used in each case. For PNe, the peak is broader than for
either the ISM or WR emission coefficients. The chemical abundances of
PNe are the result of the various nucleosynthesis processes that
modify the stellar surface chemical composition during the previous
AGB-TP phase. The exact details depend on both initial stellar mass
and metallicity of the gas from which the star formed. Individual
nebulae can be carbon-rich (C/O $>1$) or oxygen-rich (C/O $<1$)
depending on whether third dredge-up or hot bottom burning is
important in the latter stages of AGB evolution
\citep{GarciaHernandez2016}. Nitrogen can also be enhanced. Our
generic PN abundances given in the Appendix represent a carbon-rich
object with enhanced nitrogen.  The major contributors to the emission
coefficient around temperatures of $\sim 10^6$~K are bound-bound
processes (collisionally excited line emission) of these elements.

In Paper~I we showed that at early times the shocked fast wind is the
main contributor to the X-ray emission but at later times the nebular
gas is the dominant contributor, through the turbulent mixing layer or
conduction layer. Thus we could expect evolution in the abundances of
the X-ray emitting gas, from CSPN abundances through to PN
abundances. The central stars of PNe have been broadly classified into
hydrogen-rich and hydrogen-poor \citep{Mendez1991,
  Weidmann2015}. However, there is a large variety of spectral
classifications given the peculiar surface abundances and the presence
of strong stellar winds.

Although the CHANPLANS program has indicated that up to 30\% of their
sample have been detected in diffuse X-rays
\citep{Kastner2012,Freeman2014}, only three observations have
sufficient spectral resolution and count rate to enable identification
of spectral lines. Other PNe with diffuse X-ray detections in the
CHANPLANS sample have too few counts ($<200$) to make any conclusive
statements regarding spectral features (see Paper~I).

The best example is BD$+$30$^{\circ}$3639, which was observed with the
\textit{Chandra} X-ray telescope, resulting in a spectrum of $\sim
4500$ counts dominated by H-like lines of O\,{\sc viii} and C\,{\sc
  vi} and He-like lines of Ne\,{\sc ix} and O\,{\sc vii}
\citep{Yu2009}. Spectral fitting suggests that C and Ne are highly
enhanced with respect to O, compared to solar, while N and Fe are
depleted. These abundances are consistent with those of the central
star, which is a carbon-rich [WR]-type object, thus, no significant
mixing has taken place. Spectral modelling suggests a range of plasma
temperatures between $1.7\times 10^6$~K and $2.9\times 10^6$~K, which
was determined using a two-component {\it apec} model. The current fast wind
velocity is $700$~km~s$^{-1}$ and the dynamical age is only 700~yrs
\citep{Leuenhagen1996}.  This is a very young, X-ray luminous
object. Indeed, the emission coefficient curve for these abundances is
almost an order of magnitude higher than the standard PN abundance
set. We suggest that the success of the two-component model reflects
that the emission coefficient curve peaks at $\log T = 6.1$, while the
DEM for such a young object will resemble the first profile shown in
Figure~\ref{fig:dem-pn15}, that is, almost flat but with a peak close
to the maximum temperature, which is determined by the current fast
wind velocity. The range of temperatures is produced principally by
the evolution of the fast-wind velocity.

Another well-observed PN in diffuse X-rays is NGC~6543, the Cat's Eye
nebula. The spectrum of $\sim 1950$ counts has a clearly identifiable
He-like emission line triplet of O\,{\sc vii} but other lines cannot
be unambiguously identified \citep{Guerrero2015}. The central star is
hydrogen-rich, classified as OfWR(H) \citep{Weidmann2015}. Spectral
modelling with a single temperature plasma suggests a temperature of
$1.7\times 10^6$~K for the hot gas and there is no evidence for
over-abundances of C and Ne with respect ot O, which are modelled with
nebular abundances. The data quality is not good enough to determine
whether the nitrogen-to-oxygen ratio resembles that of the stellar
wind or the nebular value. The fast-wind velocity of the CSPN is
$1400$~km~s$^{-1}$ \citep{Prinja2007} and so this object is expected
to be older than BD$+$30$^{\circ}$3639, with a more evolved DEM
profile (i.e. a steeper slope) and a greater contribution to the X-ray
emission from the nebular gas through hydrodynamic interactions and/or
thermal conduction (see Paper I).

A third PN, NGC\,7027, has been detected in diffuse X-rays with 245
counts \citep{Kastner2001}. Spectral modelling indicates a plasma
temperature of $3\times 10^6$~K, near-solar abundances of O and Ne,
and overabundances of He, C, N, Mg and Si. The dynamical age of the
nebula is only $\sim 600$~yr \citep{Masson1989}, which is consistent
with the extremely high effective temperature of the central object
($T_\mathrm{eff} = 118$~kK), which is known to be rapidly evolving
\citep{Ziljstra2008}. The central star is difficult to observe because
of the low contrast with respect to the nebular emission, and no
stellar wind parameters are reported for it in the
literature. However, we can speculate that the wind will be similiar
to or even less evolved than that of BD$+$30$^{\circ}$3639. Thus, a
flat DEM profile dominated by stellar wind material with a peak around
the temperature given by the maximum wind velocity would be consistent
with the reported diffuse X-ray temperature.

Other cases that exhibit diffuse X-ray emission are the
hydrogen-deficient PNe A\,30 and A\,78 \citep[][]{Chu1995,
  Chu1997,Guerrero2012,Toala2015}. A\,30 and A\,78 are part of the
selected small number of PNe classified as born-again PNe in which the
central star has experienced a {\it very late thermal pulse}
\citep[e.g.,][]{Herwig1999}. These objects represent the coolest among
all diffuse X-ray-emitters with plasma temperatures of
$T_\mathrm{X}\lesssim$10$^{6}$~K. In these cases the carbon-rich fast
wind ($v_{\infty}\gtrsim$3000~km~s$^{-1}$) from the central star slams
into the hydrogen-poor material ejected from the very late thermal
pulse producing a variety of complex dynamical processes
(mass-loading, ablation, hydrodynamical mixing, etc) along with
photoevaporation caused by the strong ionizing UV flux from the
central star \citep[][]{Fang2014}. These processes mix the metal-rich
material creating pockets of X-ray-emitting gas, which will have an
emission coefficient similar to that of BD$+$30$^{\circ}$3639 (see
Fig.~1 central panel), enhancing the production of very soft
temperatures.  Detailed simulations on the formation and X-ray
emission from these objects will be desirable in order to deepen our
understanding on the production of X-ray-emitting gas in such unique
environments (Toal\'{a} \& Arthur in prep.).

\subsection{Wolf-Rayet Nebulae}
The abundances for NGC~6888 given in Table~\ref{tab:abundances} come
from the most recent deep spectral study of this object by
\citet{Esteban2016}. There is enhanced He and N in this object, normal
C abundance and some evidence of O deficiency, compared to
solar. Interestingly, the nebular Fe abundance is closer to solar than
to Galactic ISM abundances\footnote{$12 + \log \mbox{(Fe/H)}$ is 6.84
  (NGC~6888), 7.50 (Solar) and 5.80 (Galactic ISM),
  respectively.}. This suggests that a substantial fraction of Fe is
in the gas phase rather than being locked up in grains. It is this Fe
abundance that is responsible for the plateau seen in the emission
coefficient curve around log$_{10}(T) \sim$ 6.7. The stellar surface
abundances of WR136, as determined from stellar atmosphere model fits
to optical and UV spectra \citep{ReyesPerez2015} show increased He/O,
N/O and Fe/O abundances, while C/O is diminished.

Diffuse X-ray emission has been detected in four WR nebulae: S~308,
NGC~2359, NGC~3199 and NGC6888 around WR\,6, WR\,7, WR\,18 and
WR\,136, respectively
\citep{Toala2012,Toala2015b,Toala2016a,Toala2017}. The spectral type
of the central star of NGC~6888, WR136, is WN6, while the other three
central stars are all WN4. The stellar wind velocities of the central
stars are all very similar, between 1600 and 1800~km~s$^{-1}$.  These
WR nebulae are much more extended than PNe detected with CHANPLANS,
ranging from a radius of 2.5~pc (NGC\,2359) to 9~pc (S\,308) and a
large number of total counts ($\sim 10,000$) is required to restrict
the X-ray-emitting gas parameters. A general property of WR nebula
X-ray spectra is the presence of two distinct temperature
components. In the best-observed objects, NGC~3199 and NGC~6888, the
principal component corresponds to gas at about $1.3\times 10^6$~K and
accounts for around 90\% of the flux, while the second corresponds to
higher temperature gas of about $8.3\times10^6$~K and represents
$<8\%$ of the flux. The other two objects, S~308 and NGC~2359, have
higher temperature second components ($>10^7$~K) but the derivation is
more uncertain.

Our test simulation, reported in Appendix~\ref{sec:appwr} shows that
the two-component temperature distribution is due principally to the
abundance set in the nebular gas as long as a fully populated DEM
profile is present. The negative slope of the DEM profile means that
the higher temperature component will never dominate the emission. The
DEM profile (see Fig.~\ref{fig:wrsimdem}) spans a full range of
temperatures from photoionized gas through to the immediate post-shock
temperature of the fast wind material. This is because the complex
hydrodynamic interactions around the clumps and filaments formed by
instabilities in the wind-wind interaction raise the temperature of
the nebular material. Thus, rather than indicating two independent
components with different temperatures, the X-ray spectra are telling
us something about the gas-phase abundances in these
objects. \citet{Esteban2016} report iron abundances an order of
magnitude higher than ISM or average PN values for both S308 and
NGC~6888. The high iron abundance responsible for the second component
could be related to the life-cycle of grains in the RSG and WR stages
of evolution. There are two possibilities, (i) that silicate grains
never form in the atmosphere of the RSG (or yellow hypergiant) stage
prior to the WR stage (ii) that the grains are destroyed by shock
waves once the fast wind starts to sweep up the circumstellar medium.

\subsection{Bubbles around hot main-sequence stars}

It would be reasonable to think that hot main-sequence stars, sources
of ionizing photons and possessing fast stellar winds, are prime
candidates for diffuse X-ray emission. Indeed, many star-forming
regions have been observed with the \textit{Chandra} X-ray telescope
\citep[see, e.g.,][for a full catalogue of
  observations]{Townsley2014}. However, the data analysis of these
observations is notoriously complicated, with background subtraction
and point source excision requiring special treatment
\citep{Broos2010,Townsley2011b}. Diffuse X-ray emission has been found
associated with every massive star-forming region observed with
\textit{Chandra} and this emission has been attributed to hot plasma
from stellar wind shocks \citep{Townsley2014}. When compared to
observations at other wavelengths of the same regions, the diffuse
X-ray emission in the 0.5--7~keV band is anticorrelated with the
\textit{Spitzer} mid-IR emission tracing the boundaries of the massive
star-forming regions. It occupies the cavities in the molecular clouds
and also appears to leak out through fissures in the clouds
\citep[cf. ][]{Rogers2013}.

The massive star-forming regions detected in diffuse X-rays typically
contain clusters of massive stars and some of the regions will
certainly have had supernova activity, which will contribute to the
diffuse emission. The diffuse X-ray emission has been characterized
for a handful of objects, with the best determinations made for the
Carina Nebula ($\sim1.0\times10^6$~net counts) and M17 ($1.5\times
10^5$~net counts). The spectra were fitted with several components,
where the principal component for the Carina Nebula has $kT=0.31$~keV
($3.6\times 10^6$~K), while that for M17 has $kT = 0.28$~keV
($3.25\times 10^6$~K) \citep{Townsley2011b}. Carina is a star-forming
complex containing 8 open clusters and of order 70 ionizing sources,
including the luminous blue variable $\eta$~Car. The detection of
neutron stars in Carina implies that there have been core-collapse
supernova events. On the other hand, M17 is a very young giant HII
region ($\sim 0.5\times10^6$~yrs) with 14 known O~stars. The estimated
total band intrinsic luminosity of the Carina Nebula is an order of
magnitude higher than that of M17 ($L_\mathrm{X} = 1.7\times
10^{35}$~erg~s$^{-1}$ against $L_\mathrm{X} = 2\times
10^{34}$~erg~s$^{-1}$ ).

The Orion Nebula is a young (age $< 2\times10^6$~yrs) compact HII
region with 2 O stars and 7 early B stars. Diffuse X-ray emission from
what is known as the Extended Orion Nebula, observed with
\textit{XMM-Newton} and reported by \citet{Gudel2008}, is characterized
by a temperature $\sim 2\times 10^6$~K and an estimated total band
intrinsic luminosity 400 times less than that of M17. The diffuse
emission is offset from the position of the ionizing stars, being
located to the southwest in the low-density region furthest away from
the main ionizing front. 

Most recently, faint diffuse thermal X-ray emission characterized by a
temperature of $2\times 10^6$~K has been detected around the runaway O
star $\zeta$~Oph \citep{Toala2016b} with a total luminosity in the
0.4--4.0~keV energy range of $L_\mathrm{X} =
7.6\times10^{29}$~erg~s$^{-1}$. The diffuse emission comes from the
wake region of the bowshock structure around this isolated O9.5 star,
which is moving through the interstellar medium at
$\sim~30$~km~s$^{-1}$. This is consistent with the results of
numerical simulations of stellar wind bubbles around moving O stars by
\citet{Mackey2015}, who predicted that soft, faint X-ray emission is
produced at the wake of the bow shock in the turbulent mixing layer
between the hot, shocked stellar wind and the warm, photoionized gas
of the HII region. However, currently $\zeta$~Oph is the only isolated
O star with associated diffuse thermal X-ray emission.

We suspect that the surprisingly similar derived X-ray temperatures
for these disparate objects have a common origin, namely that
turbulent dissipation of the hot shocked fast wind energy in a mixing
layer at the interface with the photoionized gas of the HII region
forms the characteristic negative slope of the DEM profile. The
emission coefficient applied to this DEM ``selects'' the gas with
temperature close to the peak and this is representative of the
derived X-ray temperature. For typical ISM or HII region abundances
this temperature corresponds to $\log T = 6.5$, i.e. $T \sim 2\times
10^6$~K. The turbulence at the interface can be triggered by many
factors \citep{Breitschwerdt1988,Kahn1990,Mackey2015}: for single star
bubbles in uniform media, the cooling swept-up shell can become
unstable and the shadowing instability can enhance the instability and
turbulence is generated as the hot shocked wind flows through gaps in
the swept-up shell. Large-scale density gradients exist in many
observed HII regions and these lead to shear flows, which generate
turbulence \citep{Arthur2006}. Moving stars in uniform media or
density gradients also generate shear flows
\citep{Mackey2015,Arthur2006}. On the other hand, massive star-forming
regions form in clumpy molecular clouds, where density inhomogeneities
quickly lead to non-radial motions even in the photoionized gas
\citep{Mellema2006,Arthur2011,Medina2014}.

The size of the turbulent mixing layer will depend on to what extent
the stellar wind  or the HII region dominates
the dynamics of the bubble. Different authors have derived criteria
for when the stellar wind can trap the ionization front
\citep{Raga2012,Capriotti2001,Weaver1977} but essentially high stellar
wind mass-loss rate favours wind dominance and high photoionization
rate favours HII region dominance. When the photoionized gas dominates
the dynamics, a thick HII region forms outside the hot bubble and the
turbulence is suppressed. In general, the stellar wind may
dominate at very early times but the HII region can take over at later
times. More extensive
mixing layers appear to be favoured when the stellar wind is more
dominant or when the shear layer becomes well developed in bubbles
formed in density gradients (i.e., steep gradients) or around moving stars. The
luminosity of the diffuse X-ray emission will be determined by the
amount of gas in the turbulent mixing layer but the temperature is
determined by the slope of the DEM distribution for temperatures
around the peak of the emission coefficient. 

\subsection{Numerical Effects} 
Thus far we have not mentioned numerical effects in the hydrodynamic
simulations. The hydrodynamical codes providing the results for this
paper discretize the conservation form of the Euler equations and use
a piecewise linear approximation to the fluid variables to construct
the fluxes at the interfaces between computational cells by solving
the Riemann problems here \citep{Godunov1959,vanLeer1977}. The codes
are second order in time and space (see \citealp{Arthur2006,Toala2011}
for details of the codes). The numerical approximations have an
associated truncation error, which leads to numerical diffusion (i.e.,
smearing) of features in the flow. The effects of numerical diffusion
are most obvious at contact discontinuities between two fluids of
different densities but equal pressures, where diffusion reduces the
sharpness of the contact discontinuity. Furthermore, averaging of
fluid properties within grid cells blurs local, i.e., small-scale
discontinuities. Increasing the spatial resolution of the simulation
can ameliorate the effects of numerical diffusion to a large extent,
at the expense of computational memory and calculation time
requirements. The effects of numerical diffusion on the outcomes of
Eulerian hydrodynamical simulations of advected contact
discontinuities and the growth and development of Kelvin-Helmholtz
instabilities were discussed in depth by \citet{Robertson2010}.

An associated problem has been termed numerical conduction by
\citet{Parkin2010}. At a contact discontinuity, the pressure is
continuous but the density is discontinuous and so there should be a
sharp discontinuity in the temperature on either side of the
interface. Any smearing of the density distribution leads to a
corresponding smearing of the temperature distribution. This can
result in anomalous cooling in these intermediate-temperature grid
cells or even order-of-magnitude overestimates of the X-ray luminosity
\citep{Parkin2010}.

In the present paper, we have discussed the results of previously
published simulations of WR nebulae, PNe and HII regions. In the
wind-blown bubble calculations, the interaction of the fast stellar
wind with the surrounding circumstellar medium triggers instabilities
in the thin, dense swept-up shell, which are exacerbated by the
shadowing instability caused by photoionization and result in the
break-up of the shell into clumps
\citep[][Paper~I]{Toala2011,Toala2014,Arthur2015}. The clumps travel
outwards more slowly than the main shock wave and the flow of the
shocked fast wind around the clumps generates shear flows, which
ablate the clump material and form the turbulent mixing layers. In the
case of the HII region simulations \citep{Arthur2006},
Kelvin-Helmholtz instabilities are generated near the head of the
champagne-flow or bow-shock models and propagate along the interface
between the shocked stellar wind and the photoionized medium, which
leads to turbulent mixing layers downstream. In all of these
simulations, numerical resolution plays a r\^{o}le in determining some
of the properties of the clumps resulting from the instabilities, for
example, their size and density \citep{Arthur2006}. The size of the
wind injection region can influence the number of clumps formed
\citep{ToalaArthur2014}. On the other hand, short wavelength
instabilities in the shear flows are damped in our simulations because
the photoionization softens the density distribution around the dense
clumps through photoevaporation, while the timescales of the
simulations are too short for radiative cooling to be important in any
component of the gas.
  
If we compare the simulations with and without thermal conduction, for
example the PNe simulations reported in \citet{ToalaArthur2014} and
Paper~I, we can assess how extreme diffusion affects the results: it
increases the X-ray luminosity in the \textit{Chandra} soft band by an
order of magnitude. We can infer from this that numerical diffusion
will affect the X-ray luminosity of the simulated objects.  However,
the main result of the present paper is unchanged, that is, the X-ray
\textit{temperature} of the hot, diffuse gas is in the observed narrow
range [1--3]$\times10^{6}$~K as a direct consequence of the sharply
peaked emission coefficient.

In future work we intend to examine the effects of numerical
resolution and diffusion on the properties of X-ray-emitting turbulent
mixing layers in the context of the diffuse nebulae discussed in this
paper. It is also important to understand how the results of
3-dimensional simulations might differ from the 2-dimensional
simulations reported to date (see, e.g., \citealp{VanMarle2011}).

\section{SUMMARY}
In this paper, we have analyzed the plasma temperature of the hot gas
resulting from numerical simulations of diffuse nebulae around
main-sequence and evolved stars and we have compared our findings with
observations of diffuse X-rays in PNe, WR nebulae, massive
star-forming regions and wind-blown bubbles around main-sequence
stars. Our method first constructs the differential emission measure
(DEM) profile of the simulation and then combines this with the X-ray
emission coefficent $\epsilon(T)$ calculated using the atomic database
{\sc chianti}. We have studied the effect of varying the chemical
abundances on the emission-coefficient-weighted mean temperature and
on the diffuse X-ray spectrum in the \textit{Chandra} soft energy
band.  Our main conclusions are:

\begin{itemize}

\item Different abundance sets produce their own signature in the
  X-ray emission coefficient $\epsilon(T)$ in the 0.3--2.0~keV energy
  range. For all the abundance sets used in this work, except for a
  pure hydrogen case, $\epsilon(T)$ presents two features: a narrow
  peak around [1--3]$\times10^{6}$~K and a broad bump at higher
  temperatures peaking at $\log_{10}(T)\sim7.5$. The relative
  contribution of the narrow peak decreases with decreasing
  metallicity, such that the pure hydrogen case does not possess a
  narrow peak. The absolute value of $\epsilon(T)$ is directly related
  to the X-ray luminosity and increases with increasing metallicity.

\item The emission coefficient acts as a filter for the differential
  emission measure (DEM) profile of the gas and emphasizes the
  contribution of gas with temperatures close to that of the peak of
  $\epsilon(T)$ even though the actual temperature distribution of the
  source is much more complex. For low metallicity gas, the peak of
  the emission coefficient occurs at slightly higher temperatures.

\item The DEM profiles calculated from numerical simulations of PNe,
  WR nebulae and stellar wind bubbles in H\,{\sc ii} regions around
  main-sequence stars are all remarkably similar and have steep
  negative slopes over the temperature range $5 < \log_{10}(T) <
  8$. The emission-coefficient-weighted mean temperature of these
  objects is thus always close to the temperature of the peak of the
  emission coefficent in the X-ray band. We showed that the averaged
  plasma temperature is a good estimate to the X-ray-emitting
  temperature obtained from spectral fits, that is, $T_\mathrm{A}
  \approx T_\mathrm{X}$. This is the reason why the temperature
  derived from observations always turns out to be in the range
  [1--3]$\times10^6$~K.

\item The DEM profiles of the simulated diffuse nebulae are all so
  similar because in each case turbulent mixing layers transfer energy
  from the hot shocked stellar wind to the warm ($T \sim 10^4$~K)
  photoionized gas, where it is dissipated. It does not matter whether
  thermal conduction is included in the simulations or not. Turbulent
  mixing layers are only produced in 2D or 3D simulations of diffuse
  nebulae and should be studied in more detail.

\item The second temperature component obtained from spectral analysis
  of the diffuse X-ray emission from WR nebulae is due in part to the
  shocked fast wind material from the central star. Fast stellar winds
  from WR stars have much higher mass-loss rates than for either
  main-sequence stars or the central stars of PNe. A second
  temperature component has not been reported for these other sorts of
  objects. In WR nebulae, the wind contributes significantly to the
  spectrum in the $\sim$0.8--1.2~keV range.
  
\item A diffuse nebula in the SMC will appear to be slightly hotter
  and up to an order of magnitude less luminous in X-rays than the
  same object in the Milky Way, simply on the basis of the emission
  coefficient in the soft X-ray band (0.3--2.0~keV).
  
\end{itemize}

Finally, we would like to remark that the number of PNe and WR nebulae
that exhibit diffuse X-ray emission is realatively small (4 WR nebulae
and only $\sim$30\% of all PNe observed in the CHANPLANS survey) and
in some cases the spectral quality is not good enough to make a confident
analysis of the spectral properties of the X-ray-emitting plasma. It
would be interesting to use the future generation of X-ray satellites
\citep[e.g., $Athena+$;][]{Nandra2013} to try to unveil the presence
of hotter components in the diffuse emission and properly characterise
the abundances of the hot gas in these objects through resolved line
emission.

\section*{Acknowledgements}
JAT and SJA are funded by UNAM DGAPA PAPIIT projects IA100318 and
IN112816, respectively. This work has made extensive use of NASA's
Astrophysics Data System. We thank the referee for pointing us to the
outstanding work of \citet{Strickland1998}.

\appendix

\section{Abundance sets}
\label{sec:appab} 
Here we present the details of the abundances used for our calculations.

\begin{table*}
\begin{center}
 \caption{Abundance sets.}
 \label{tab:abundances}
 \begin{tabular}{|l|ccc|cccc|cc|}
  \hline
  \multicolumn{1}{l|}{Element} & \multicolumn{3}{|c|}{ISM} & \multicolumn{4}{|c|}{PNe}  & \multicolumn{2}{|c}{WR nebulae}\\
           &  MW$^\mathrm{a}$   & LMC$^\mathrm{b}$   & SMC$^\mathrm{c}$    & MW$^\mathrm{a}$    & LMC$^\mathrm{d}$   & SMC$^\mathrm{e}$  & BD$+$30$^{\circ}$3639$^\mathrm{f}$  & NGC\,6888$^\mathrm{g}$ & BAT99-11$^\mathrm{h}$\\
  \hline
  He      & 10.99 & 10.95 & 10.91  & 11.00 & 11.10 & 10.99 &  10.99                & 11.20     & 11.01 \\
  C       & 8.40  & 7.90  & 7.40   & 8.89  & 8.52  & 7.41  &  10.59                & 8.50      & 7.90  \\
  N       & 7.90  & 6.90  & 6.50   & 8.26  & 8.24  & 7.35  &  7.90                 & 8.53      & 6.56  \\
  O       & 8.50  & 8.40  & 8.00   & 8.64  & 8.42  & 7.89  &  9.53                 & 8.49      & 8.23  \\
  Ne      & 8.09  & 7.39  & 7.39   & 8.04  & 7.60  & 7.14  &  8.09                 & 7.81      & 6.83  \\
  Na      & 5.50  & 5.50  & 5.50   & 6.28  & 5.50  & 5.50  &  6.28                 & 5.50      & 5.50  \\
  Mg      & 7.10  & 6.80  & 6.40   & 6.20  & 6.15  & 6.40  &  6.20                 & 7.10      & 6.80  \\
  Al      & 4.90  & 4.60  & 4.20   & 5.43  & 4.60  & 4.20  &  5.43                 & 4.90      & 4.60  \\
  Si      & 6.50  & 6.70  & 6.30   & 7.00  & 7.10  & 6.30  &  7.00                 & 6.50      & 6.70  \\
  S       & 7.51  & 6.77  & 6.77   & 7.00  & 6.94  & 6.98  &  7.00                 & 7.23      & 6.20  \\
  Fe      & 5.80  & 5.49  & 5.10   & 5.70  & 5.49  & 5.10  &  5.70                 & 6.84      & 5.49  \\
  \hline
 \end{tabular}
\begin{tablenotes}
 \item $^\mathrm{a}$Predefined for Cloudy \citep[][]{Ferland2013},
   $^\mathrm{b}$\citet{Schenck2016}, \citet{Hughes1998} and
   \citet{Korn2000}, $^\mathrm{c}$\citet{Korn2000},
   $^\mathrm{d}$\citet{Dopita1997}, $^\mathrm{e}$\citet{Idiart2007},
   $^\mathrm{f}$\citet{Marcolino2007} and \citet{Yu2009},
   $^\mathrm{g}$\citet{Esteban2016}, and
   $^\mathrm{h}$\citet{Stock2011}.
\end{tablenotes}
\end{center}
\end{table*}

\section{Simulation of a Wolf-Rayet Nebula}
\label{sec:appwr}

\begin{figure*}
\begin{center}
\includegraphics[height=0.5\linewidth]{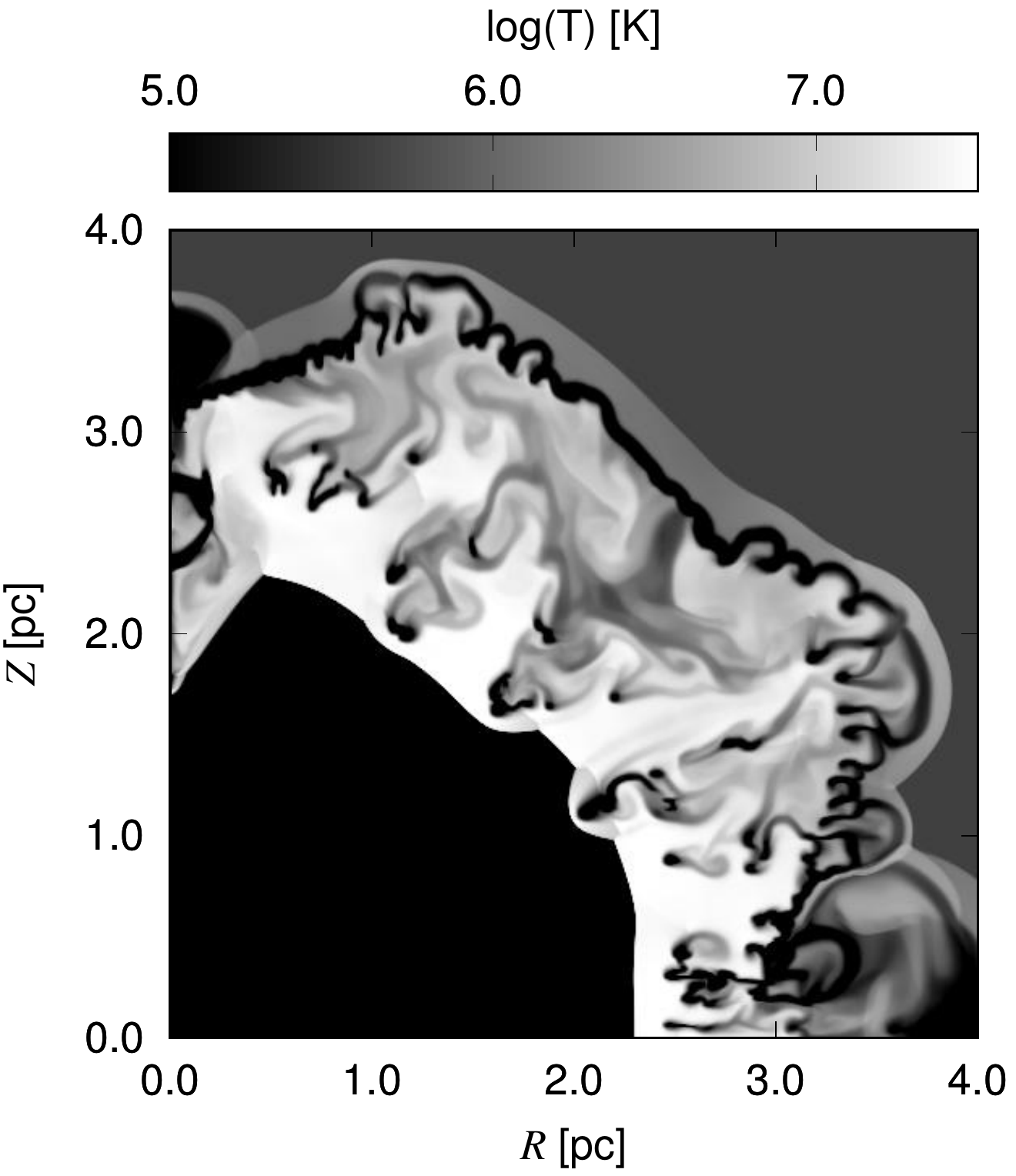}~
\includegraphics[height=0.5\linewidth]{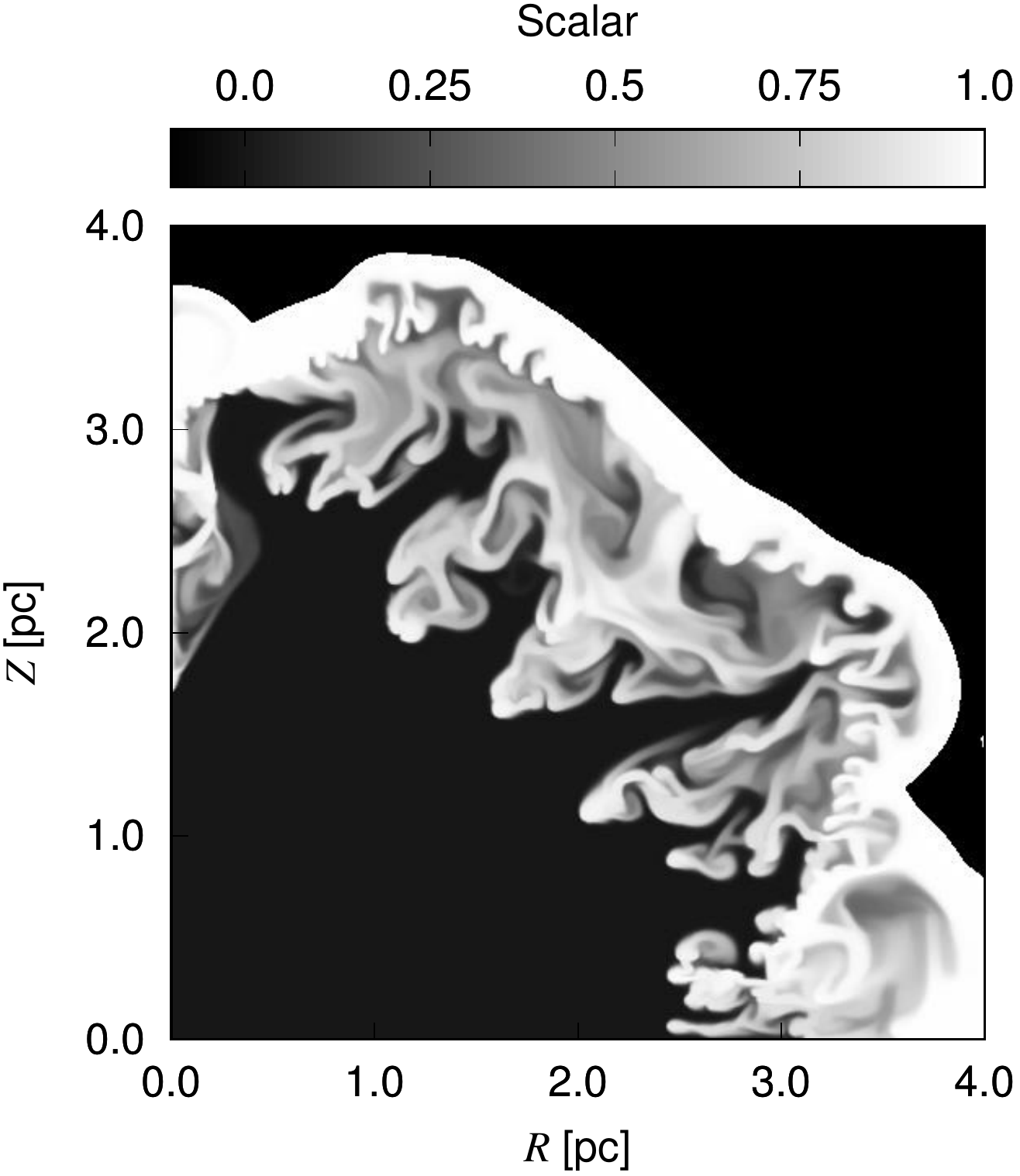}
\end{center}
\caption{Simulation of a generic WR nebula
  \protect\citep[see][]{Arthur2015}. Left panel: gas temperature above
  $10^5$~K. The corrugated swept-up RSG wind shell (black) and the
  hot, shocked main-sequence bubble material outside it can be clearly
  distinguished. Right panel: advected scalar, which is used to trace
  the different components of the nebula. In this panel, negative
  values correspond to unperturbed main-sequence bubble material,
  scalar value 0 corresponds to unmixed fast stellar wind, while
  nebular gas (shocked main-sequence bubble material and RSG wind) has
  scalar value 1. Intermediate values of the scalar indicate mixing
  regions between the fast wind and the nebular material.}
\label{fig:wrsim}
\end{figure*}

We have performed 2D axisymmetric simulations of the interaction
between a fast wind and a dense circumstellar medium with parameters
appropriate to a WR nebula, though without being tailored for any
object in particular \citep{Arthur2015}. The simulations were done on
a $500^2$ grid using the same code as used for the planetary nebula
simulations of \citet{ToalaArthur2014} and include photoionization and
radiative cooling but thermal conduction is turned off. The chemical
abundance set used in this model is appropriate to NGC~6888, the
nebula around the WR star WR136. These simulations assume that the
initial circumstellar medium represents the expelled RSG envelope and
is composed of $15 M_\odot$ of material that was ejected at a steady
rate over 200,000~years at 15~km~s$^{-1}$ into a uniform, hot ($T =
4.27 \times 10^5$~K), low-density ($n = 0.01$~cm$^{-3}$), ionized
medium (which represents the main-sequence hot, shocked stellar wind
bubble). The WR star stellar wind is taken to have a constant velocity
$v_{\infty} = 1600$~km~s$^{-1}$ and mass-loss rate $\dot{M} = 7\times
10^{-5} M_\odot$~yr$^{-1}$. The dense shell marking the rim of the RSG
wind is initially 2.7~pc from the star.

At the time shown in Figure~\ref{fig:wrsim}, 15,000~years after the
onset of the fast wind, the RSG material has been completely swept up
and the bubble has broken out into the low density medium beyond. This
is evident in the distribution of the scalar value and the gas
temperature shown in the figure. The wind-wind interaction forms an
unstable dense shell, which breaks up into knots and filaments as it
travels down the $r^{-2}$ density gradient. Complex hydrodynamic
interactions as shock waves are refracted around the clumps lead to a
full range of gas temperatures in the wind bubble, as can be seen from
the DEM profile (Fig.~\ref{fig:wrsimdem}). Fast wind material occupies
the higher temperature range, $\log T > 7.2$, while nebular material
dominates temperatures in the range $\log T < 7.2$. The unshocked
main-sequence-bubble material has a negligible ($<1\%$) contribution
to the DEM in the temperature bin at $\log T = 5.63$ because the
density is so low in this gas.




\end{document}